\documentclass{aa}

\usepackage{graphicx,color}
\usepackage{txfonts}
\usepackage{natbib}
\bibpunct{(}{)}{;}{a}{}{,} 

\begin{document}

   \title{PIC simulation study of the interaction between a relativistically moving leptonic micro-cloud and ambient electrons}

   \titlerunning{Instabilities driven by a lepton cloud}
   \author{M. E. Dieckmann
          \inst{1}
          \and
          G. Sarri\inst{2}
          \and
          S. Markoff\inst{3}
          \and
          M. Borghesi\inst{2}
          \and
          M. Zepf\inst{2}
          }

   \institute{Department of Science and Technology (ITN), Linkoping University, 60174 Norrkoping, Sweden\\
              \email{mark.e.dieckmann@liu.se}
         \and
              Centre for Plasma Physics, Queen's University Belfast, BT7 1NN, Belfast, United Kingdom\\
             \email{g.sarri@qub.ac.uk \& m.zepf@qub.ac.uk}
         \and
             Anton Pannekoek Institute for Astronomy/GRAPPA,
             University of Amsterdam, 1098 XH, Amsterdam, The Netherlands\\
             \email{s.b.markoff@uva.nl}
             }

\date{}

\abstract 
   {The jets of compact accreting objects are composed of electrons and a mixture of positrons and ions. These outflows impinge on the interstellar or intergalactic medium and both plasmas interact via collisionless processes. Filamentation (beam-Weibel) instabilities give rise to the growth of strong electromagnetic fields. These fields thermalize the interpenetrating plasmas.} 
   {Hitherto, the effects imposed by a spatial non-uniformity on filamentation instabilities have remained unexplored. We examine the interaction between spatially uniform background electrons and a minuscule cloud of electrons and positrons. The cloud size is comparable to that created in recent laboratory experiments and such clouds may exist close to internal and external shocks of leptonic jets. The purpose of our study is to determine the prevalent instabilities, their ability to generate electromagnetic fields and the mechanism, by which the lepton micro-cloud transfers energy to the background plasma.}
{A square micro-cloud of equally dense electrons and positrons impinges in our particle-in-cell (PIC) simulation on a spatially uniform plasma at rest. The latter consists of electrons with a temperature of 1 keV and immobile ions. The initially charge- and current neutral micro-cloud has a temperature of 100 keV and a side length of 2.5 plasma skin depths of the micro-cloud. The side length is given in the reference frame of the background plasma. The mean speed of the micro-cloud corresponds to a relativistic factor of 15, which is relevant for laboratory experiments and for relativistic astrophysical outflows. The spatial distributions of the leptons and of the electromagnetic fields are examined at several times.}
   {A filamentation instability develops between the magnetic field carried by the micro-cloud and the background electrons. The electromagnetic fields, which grow from noise levels, redistribute the electrons and positrons within the cloud, which boosts the peak magnetic field amplitude. The current density and the moduli of the electromagnetic fields grow aperiodically in time and steadily along the direction that is anti-parallel to the cloud's velocity vector. The micro-cloud remains conjoined during the simulation. The instability induces an electrostatic wakefield in the background plasma.}
   {Relativistic clouds of leptons can generate and amplify magnetic fields even if they have a microscopic size, which implies that the underlying processes can be studied in the laboratory. The interaction of the localized magnetic field and high-energy leptons will give rise to synchrotron jitter radiation. The wakefield in the background plasma dissipates the kinetic energy of the lepton cloud. Even the fastest lepton micro-clouds can be slowed down by this collisionless mechanism. Moderately fast charge- and current neutralized lepton micro--clouds will deposit their energy close to relativistic shocks and hence they do not constitute an energy loss mechanism for the shock.}


  \maketitle

\section{Introduction}

Accreting compact objects (neutron stars and black holes) can emit
relativistic plasma jets
\citep[e.g.][]{Lovelace1976,BlandfordZnajek1977,BlandfordPayne1982}. Some
recent examples of such jets are those inferred for gamma-ray bursts
(GRBs) \citep{Cenko10}, or now directly observed at near-Event Horizon
scales in active galactic nuclei (AGN; \citealt{Hada11,Doeleman12})
and X-ray binaries (XRBs)
\citep[e.g.][]{Mirabeletal1992,Fender01,Corbel02}. Energetic
processes, which yield a heating and subsequent thermalization of the
plasma via the interaction of charged particles with the
self-generated electromagnetic fields, can be expected to develop
within the jet and at its collision boundary with the ambient
interstellar or intergalactic medium. The large mean free path of the
particles in the jet and in the ambient medium implies that these
processes can not be mediated by binary collisions between particles;
rather relativistic beam instabilities and collisionless shocks will
thermalize the plasma flow.  Such shocks can be external, forming
between the jet and the ambient medium, or internal, resulting from
strong spatial variations of the plasma's mean flow velocity within
the jet.

The huge energy density carried by a relativistic jet in the form of
radiation, electromagnetic fields and kinetic energy can trigger
pair-production. It is thus likely that such jets carry a significant
fraction of positrons, that can react more easily to electromagnetic
fields than heavier ions. We may thus neglect in some cases the ion
contribution to the plasma processes close to the external and
internal shocks; for example if the processes take place on short
spatio-temporal scales. More specifically, the ion reaction is
negligible if the instabilities grow and saturate on time scales that
are comparable to a few tens of inverse plasma frequencies and if the
spatial scales of the plasma structures are of the order of an
electron skin depth. These are millisecond- and kilometre scales if
the electron number density is $\approx$ 1 cm$^{-3}$.

Accordingly, a wide range of theoretical
\citep{Medvedev99,Brainerd00,Bret06,Milos06,Bret10,Tautz12} and
particle-in-cell (PIC) simulation studies
\citep{Sakai00,Honda00,Silva03,Jaroschek04,Medvedev05,Dieckmann09a,Dieckmann09b,Vieira12}
have addressed the instabilities of counterstreaming lepton beams and
the formation of shocks in pair plasmas
\citep{Kazimura98,Haruki03,Sironi11,Sironi13,Bret13} that are
triggered when plasmas collide at a relativistic speed.


In the meantime, recent advances in laser technology have opened up
another means to study the processes expected to take place in
relativistic leptonic flows. It is now possible to produce clouds of
electrons and positrons in the laboratory that are large and dense
enough to initiate their collective (plasma) behaviour. High-energy
electron/positron clouds have been reported by
\citet{Chen10}. However, the small percentage of positrons in the
leptonic cloud (~10 \%) and the relatively low density (transverse
size smaller than the collisionless skin depth) prevent plasma
behavior to occur. Using this experimental approach, these
characteristics can be achieved with the 10 PW laser facilities that
will be available at the end of this decade
\citep{Ridgers12}. Adopting a different laser-driven scheme first
proposed by \citet{Sarri14}, dense and neutral leptonic clouds have
been recently achieved \citep{Sarri13}. The small divergence and
neutrality of the reported beams allow for the study of the growth of
plasma instabilities, that are likely relevant for relativistic
astrophysical flows. In the following paragraphs we put forward a
scenario motivated in part by recent PIC simulations and experiments,
which we here propose to study further with a pointed PIC simulation.

A plasma shock transforms the directed flow energy of the upstream
flow into thermal energy of the downstream plasma. If the shock were
mediated by binary collisions between particles, it would form a sharp
boundary perfectly separating the two plasma populations with
different mean flow speeds, temperatures and field energy
densities. However, the collisionless nature of the plasma expected
for astrophysical jets implies that their internal and external shocks
are mediated instead by electromagnetic fields. The self-generated
electromagnetic fields are able to sustain the plasma shock and can
confine the bulk of the heated plasma in its downstream
region. However, unlike the case of collisional shocks, the
confinement will not be perfect. Energetic electrons, positrons and
ions can leak out from the downstream region and move upstream. PIC
simulations of relativistic shocks involving electrons and ions
\citep{Martins09} or electrons and positrons \citep{Chang08} show
evidence for such a leakage. The leaking particles will not
necessarily have a spatially uniform thermal distribution. Hence
micro-clouds of electrons and positrons, with dimensions comparable to
the thickness of the transition layer of a leptonic shock, can exist
close to sharp boundaries like shocks, that separate plasma
populations with vastly different mean kinetic energy densities. For
the sake of this work, we will neglect escaping ions since their large
inertia implies that they can usually not move in unison with the
escaping leptons.

Escaping clouds of relativistic electrons introduce a strong net
current ahead of the shock. These clouds will either lose their energy
to the growth of strong magnetic fields \citep{Sarri12} or they will
drive a return current in the plasma ahead of the shock. For the
latter case, two-stream instabilities will develop and thermalize the
electron clouds. If a sufficiently dense population of positrons
exists close to the shock, then the escaping electrons can be
charge-and current-neutralized by comoving positrons. This overall
charge-and current-neutralized beam can drive electromagnetic
instabilities ahead of the shock. A foreshock will then develop, which
expands upstream along the normal direction of the shock
\citep{Chang08} until the point where the leaking particles can no
longer drive plasma instabilities. In principle, because of the lack
of binary collisions, the micro-clouds could propagate indefinitely if
they did not drive plasma instabilities, in which case their energy
would be lost from the shock.  It is thus interesting to determine
whether or not a skin depth-scale leptonic cloud will in fact interact
with an ambient plasma through collective plasma instabilities or if
such clouds constitute an energy loss mechanism for a relativistic
shock.

Our particle-in-cell (PIC) simulation study presented here is thus motivated 
by this need to better understand the physics of localized relativistic 
lepton clouds. Using a two-dimensional PIC simulation, we examine the impact 
of a small electron-positron cloud on an ambient plasma consisting of 
electrons and immobile ions. The number density and the average speed of the 
initially thermal electrons and positrons is equal and spatially uniform 
within the cloud. Consequently, the lepton cloud is free of any net charge 
and net current at the start of the simulation. The relative speed between 
the cloud and the ambient plasma corresponds to a Lorentz factor of 15, which 
is likely somewhat higher than that of the jets of XRBs, but typical for AGN 
\citep[e.g.][]{Listeretal2013} . Only lepton clouds that move faster than the 
jet and the shock can move from the region behind the shock into the upstream 
region of the shock.

The temperature of the ambient electrons is 1 keV and that of the
cloud leptons is 100 keV, typical values inferred from pair
Comptonization models in, e.g., X-ray binaries \citep[see,
e.g.][]{SunyaevTitarchuk1980,EsinMcClintockNarayan1997,MerloniFabian2002},
or shock-heated ambient gas \citep[e.g.][]{Lanzetal2015} in order to
explore a physical scale that is both feasible to simulate, and also
potentially relevant for real systems.  The pair cloud is launched
initially in a vacuum and then collides with the ambient electron
plasma that is spatially bounded and uniform. We consider a normal
incidence of the pair cloud with respect to the boundary that
separates the ambient plasma from the vacuum. The slow growth of the
instability implies that all plasma processes develop in a region that
is well-separated from this boundary and the latter is thus not
important for the plasma dynamics.

This paper is structured as follows. Section 2 discusses the
simulation code and the initial conditions. Section 3 presents the
simulation results. Section 4 is the discussion.


\section{The simulation code and the initial conditions}

PIC codes employ a mixed Lagrangian-Eulerian scheme \citep{Dupree63} to solve the following set of equations. Amp\`ere's law and Faraday's law:
\begin{equation}
\mu_0 \epsilon_0 \frac{\partial }{\partial t} \mathbf{E}(\mathbf{x},t) = \nabla \times \mathbf{B}(\mathbf{x},t) - \mu_0 \mathbf{J}(\mathbf{x},t),
\label{Ampere}
\end{equation}
\begin{equation}
\frac{\partial}{\partial t} \mathbf{B}(\mathbf{x},t) = -\nabla \times \mathbf{E}(\mathbf{x},t) \label{Faraday},
\end{equation}
are solved on a numerical (Eulerian) grid. The EPOCH code \citep{Ridgers13}, which we use here, solves Gauss' law as a constraint and $\nabla \cdot \mathbf{B} = 0$ to round-off precision. 

Coulomb collisions between particles are negligible for processes within
and close to the transition layers of leptonic shocks if they develop
on time scales that are comparable to the inverse plasma frequency
$\omega_{p,i} = {(n_i e^2/\epsilon_0 m_e)}^{1/2}$, where $n_i$ is the
number density of the leptons of species $i$, $e$ is the elementary
charge and $m_e$ is the electron mass. The absence of collisions
implies that the plasma is not automatically in a thermal equilibrium,
which would be characterized by a Maxwellian velocity distribution.

A more appropriate description of a collisionless plasma is thus given
by the phase space density distribution
$f_i(\mathbf{x},\mathbf{v},t)$, for which the position $\mathbf{x}$
and the velocity $\mathbf{v}$ are independent variables. Each plasma
species $i$ is represented by a separate phase space density
distribution $f_i(\mathbf{x},\mathbf{v},t)$. The number density and
the average speed are derived from this distribution as
$n_i (\mathbf{x},t) = \int f_i(\mathbf{x},\mathbf{v},t)\, d\mathbf{v}$
and
$\hat{\mathbf{v}}_i (\mathbf{x},t) = \int \mathbf{v} \,
f_i(\mathbf{x},\mathbf{v},t)\, d\mathbf{v}$,
respectively. The charge density in Gauss' law is obtained from the
number density, and the current density in Amp\`ere's law is computed
using the average speed.

A PIC code approximates the phase space density distribution
$f_i(\mathbf{x},\mathbf{v},t)$ of each plasma species $i$ by an
ensemble of Lagrangian- or computational particles (CPs). These CPs
correspond to phase space blocks with a charge-to-mass ratio that
equals that of the plasma particles they stand for. The force imposed
on charged particles by electromagnetic fields is proportional to
their charge $q$, while their inertia is determined by their mass
$m$. The acceleration is thus proportional to $q/m$ and the plasma
evolution does not depend on any other quantity. The numerical values
of the charge $q_i$ and mass $m_i$ of a CP that represents particles
of the species $i$ can thus be different from that of individual
particles of this species as long as their ratio is the same. The
relativistic momentum $\mathbf{p}_j = m_i \Gamma_j \mathbf{v}_j$ of
each CP with index $j$ of the species $i$ is evolved in time by a
numerical approximation of the Lorentz force equation
\begin{equation}
\frac{d}{dt} \mathbf{p}_j = q_i \left ( \mathbf{E}(\mathbf{x}_j) + \mathbf{v}_j \times \mathbf{B}(\mathbf{x}_j) \right ),
\label{Lorentz}
\end{equation}
and its position $\mathbf{x}_j$ is updated through $\frac{d}{dt}\mathbf{x}_j = \mathbf{v}_j$. 

The algorithm on which explicit PIC codes are based can be summarized
as follows. The current carried by each CP is interpolated to the
grid. The summation over all current contributions gives the
macroscopic current $\mathbf{J}(\mathbf{x},t)$, which is defined on
the grid. The electromagnetic fields are updated per time step with
this macroscopic current, and then interpolated back to the position
$\mathbf{x}_j$ of each CP and the particle momentum is then updated
using the new electromagnetic fields. Each cycle advances the plasma
and the field distribution by a finite time step $\Delta_t$. A more
detailed description of the PIC method can be found elsewhere
\citep[e.g.][]{Dawson83}.

Our simulation resolves the x-y plane and all three momentum components. 
The boundary conditions are periodic along $y$ and open along $x$. We 
introduce three different species and we label their respective parameters 
by the subscripts $b, e$ and $p$. Background electrons (labelled $b$) with 
the charge $-e$ and mass $m_e$, the number density $n_b=n_0$, the plasma 
frequency $\omega_p = {(n_0 e^2/\epsilon_0 m_e)}^{1/2}$ and the skin depth
$\lambda_e = c / \omega_p$ are introduced into the simulation box. The total 
box size is given by $-1.8 \lambda_e \le x \le 117 \lambda_e$ and $-13.4 
\lambda_e \le y \le 13.4 \lambda_e$. All electric and magnetic field components 
are set to zero at the simulation's start. The electromagnetic fields are 
thereafter computed from the plasma currents.

The background electrons are placed into the interval
$0 \le x \le 115 \lambda_e$. The ambipolar electric field, which
develops after some time at the boundary between the electrons and the
vacuum bands at large and low $x$, reduces the number of electrons
that can escape through the open boundary. The background electrons
are at rest in the simulation frame and they have a Maxwellian
velocity distribution with a temperature $T_b = 1$ keV. The ions form
an immobile background of positive charges, which compensates the
negative charge of the electrons. Immobile ions are introduced
implicitly into PIC simulations if we introduce mobile electrons and
set the electric field to zero. The background electrons and the
immobile ions correspond to the ambient plasma. The ambient plasma in
laboratory experiments is residual gas, which has been ionized by
secondary x-ray radiation from the solid target, which is used for
pair production. The ambient plasma could correspond to, e.g.,
shock-heated interstellar medium ahead of astrophysical jets or a
surrounding accretion flow.  

A square lepton cloud, which consists of electrons (label $e$) and
positrons (label $p$) with equal number densities $n_e = n_p = n_0$,
is placed in the (vacuum) interval
$-0.9\lambda_e \le y \le 0.9\lambda_e$ and
$-1.8\lambda_e \le x \le 0$. The side length of the cloud is
$1.8\lambda_e$ or 2.5 in units of the total leptonic skin depth
$\lambda_e / \sqrt{2}$ of the $e^+ e^-$ cloud. The mean speed $v_c$ of
the cloud is aligned with the x-direction and yields the relativistic
factor $\Gamma_c = {(1-v_c^2/c^2)}^{-1/2} = 15$. Its momentum
distribution in the cloud's frame of reference is a relativistic
Maxwellian with the temperature $T_c = 100$ keV.

As mentioned before, the temperatures for the ambient
electrons and the lepton clouds are representative of some selected
systems, but would be too high to represent the typical interstellar
medium or jet plasma far from the base, respectively.  The high value
is chosen for numerical reasons; a high temperature increases the
Debye length $\lambda_{D} = {(\epsilon_0 k_B T_b / n_0 e^2)}^{1/2}$ of
the background electrons. This Debye length, which is the smallest
scale over which collective plasma interactions are more important
than Coulomb collisions, sets the cell size in the PIC simulation and
it is proportional to the time step. A high temperature $T_b$ speeds
up the simulation and hence it allows us to resolve a larger spatial
domain. The thermal noise levels in the plasma increase with the
plasma temperature. These fluctuations set the initial amplitude of
the unstable waves. The growth and saturation of the waves is thus
accelerated by a high temperature, which decreases further the
computational cost of the simulation. A temperature of 100 keV
corresponds to a thermal spread of the electrons, which is only a
minor fraction of the relative speed between the lepton cloud and the
ambient plasma. Thermal effects on the beam instabilities are thus
negligible and the actual value of the plasma temperature is not
important. The rapid growth of the instability that we observe
implies that it must develop close to the transition layer of the
external shock of the jet, where the interstellar medium and the jet
plasma are hot and closer to our initial values for the temperature.

The simulation box is resolved by $8000 \times 1600$ grid cells along $x$ and $y$ respectively. The cells have a uniform spacing $\Delta_x = 0.015\lambda_e$ along both directions. The background electrons are represented by a total of $1.4 \times 10^9$ CPs and the electrons and positrons of the cloud by a total of $6.1 \times 10^7$ CPs, respectively. The cloud will travel along the x-direction, which is the vertical direction in all figures.

In what follows, we will normalize the electric field as $e
\mathbf{E}/\omega_p m_e c$, the magnetic field as $e \mathbf{B} /
\omega_p m_e$, particle speeds as $\mathbf{v} / c$, the macroscopic
current (on the grid) as $\mathbf{J} / e n_o c$ and the densities as
$n_{b,e,p} (x,y) / n_0$. Space and time are normalized as $x /
\lambda_e$, $y / \lambda_e$ and $\omega_p t$. The Maxwell-Lorentz set
of equations can be normalized with these substitutions and our
results can be scaled to any density that results in a weakly
correlated plasma.

\section{Simulation results}

An unmagnetized relativistic leptonic beam of infinite extent, that
interacts with a uniform background plasma, can be unstable to the
two-stream mode, to the oblique mode and to the filamentation mode
(See the review by \citet{Bret10} and references therein). However,
the two-stream and oblique mode instabilities are suppressed in our
simulation by the tiny size of the lepton cloud. The resonance
condition implies that the longitudinal component $k_u$ of the wave
vector of two-stream and oblique modes, which is aligned with the mean
velocity vector of the cloud $\mathbf{v}_c \parallel \mathbf{x}$, is
$\omega_p / k_u \approx v_c$.  This wave vector component corresponds
with $v_c \approx c$ to the wavelength $\lambda_u = 2\pi \lambda_e$ in
the simulation frame. This resonance condition states that a
perturbation, which is moving at almost the speed of light in the
simulation frame, has to interact resonantly with the background
electrons that are at rest in this frame. The only resonance frequency
of unmagnetized electrons is the plasma frequency. The value of
$\lambda_u$ exceeds the cloud's thickness of $1.8 \lambda_e$ by a
factor of 3. The resonant instabilities can, however, only grow if the
cloud can accommodate an integer number of wave periods along the flow
velocity vector and these instabilities are thus suppressed. The same
limitation holds for laser-plasma experiments, where the longitudinal 
dimension of the beam is smaller than $\lambda_e$.

In contrast, the filamentation instability is a non-resonant
instability and obeys other constraints than the resonant two-stream
and oblique modes. The wave vector of the filamentation modes is
oriented almost perpendicularly to $\mathbf{v}_c$. The wavelength of
the filamentation modes can be much smaller than $\lambda_u$ if the
thermal speed of the leptons is small compared to the relative speed
between the ambient plasma and the cloud plasma, which is the case
considered here. The limited longitudinal extent of the cloud does not
constrain the growth of the filamentation instability, because there
is no condition on the wavelength along this direction. Our simulation
is thus designed to test if and to what extent the filamentation
instability can grow.

\subsection{Time t=27.5}

\begin{figure}
\centering
\includegraphics[width=\columnwidth]{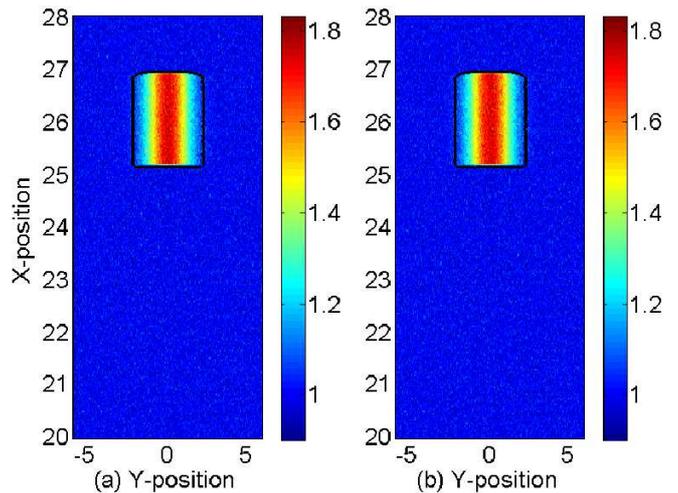}
\caption{The lepton density distribution at the time $t$=27.5. Panel (a) shows the normalized density $n_b (x,y) + n_e (x,y)$ of the background electrons and the cloud's electrons. Panel (b) shows the normalized density $n_b (x,y) + n_p (x,y)$ of the background electrons and the cloud's positrons. The color scale is linear. The black curve outlines the contour where the lepton cloud's number density is 0.06 $n_0$.}\label{Lepton149}
\end{figure}

The number density distributions of the background electrons and of
the cloud's electrons and positrons are shown in Figure
\ref{Lepton149} at the simulation time 27.5. The fastest particles of
the lepton cloud with the speed $v \lesssim c$ have propagated for a
distance of $\approx 27$ during this time. The lateral density
distribution resembles a Maxwellian, which is a consequence of the
finite evolution time and the Maxwellian momentum distribution. The
relativistic mean speed of the lepton bullet along $x$ implies that
thermal diffusion is less pronounced along this direction.
The spatial distributions $n_e (x,y)$ and $n_p (x,y)$ of the cloud's
electrons and positrons are practically identical. The cloud should be
almost charge- and current neutral and the electromagnetic fields
weak. The number density distribution $n_b (x,y)$ of the background
electrons appears to be unaffected.

\begin{figure}
\centering
\includegraphics[width=\columnwidth]{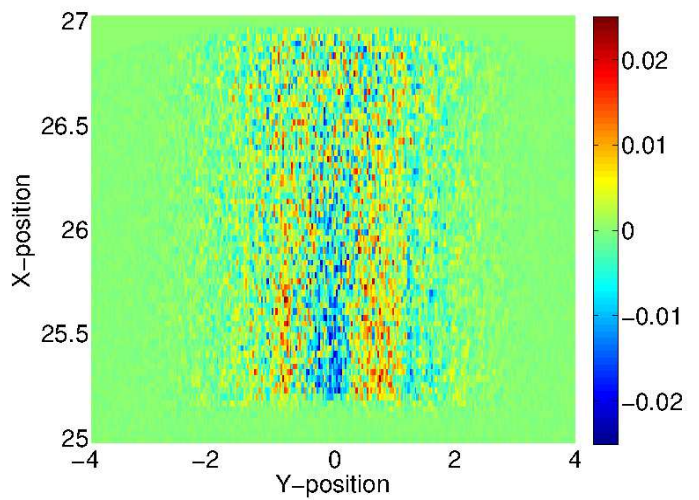}
\caption{The net charge density of the cloud at the time $t$=27.5. The net charge is calculated as $n_p (x,y) - n_e (x,y)$ (normalization to $n_0 e$) and the color scale is linear.}\label{NetCharge149}
\end{figure}

The normalized number density difference $n_d (x,y) = n_p (x,y)-n_e
(x,y)$ yields a more accurate measure of the charge and current
density distributions of the lepton cloud. Figure \ref{NetCharge149}
reveals a modulation of $n_d (x,y)$ on a larger scale.
The modulations are most pronounced at the back of the cloud at $x\approx 
25.5$. A minimum of $n_d (x,y)$ is observed at $y\approx 0$, which is flanked 
by two maxima at $y\pm 0.7$. The interaction of the electrons and positrons 
of the cloud with the
  background electrons is mediated by the micro-currents of the
  particles. The direction of the micro-current vectors of the cloud
  particles, measured in the reference frame of the background
  electrons, depends on the particle charge. The same holds for the
  direction of the force between background electrons and the
  electrons and positrons of the cloud. Consequently the latter are
  redistributed differently within the cloud by their interaction with
  the background electrons. The space charge associated with this
modulation will result in an electric field. This electric field will
be almost parallel to the y-axis; it will have a weak $E_x$-component,
since the oscillations get weaker with increasing $x$. The space
charge and the relativistic cloud speed will furthermore result in a
magnetic $B_z$ component in the laboratory frame.

\begin{figure}
\centering
\includegraphics[width=\columnwidth]{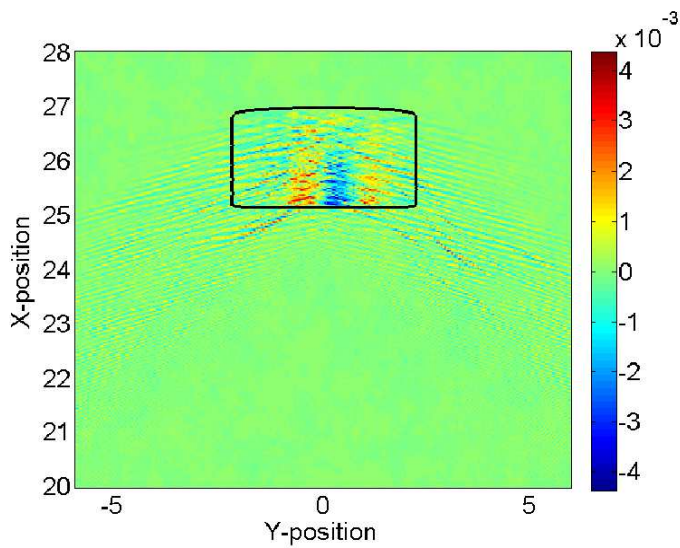}
\caption{The normalized magnetic amplitude $B_z (x,y)$ at the time $t$=27.5. The color scale is linear. The black curve outlines the contour where the lepton cloud's number density is 0.06 $n_0$.}\label{MagneticBz149}
\end{figure}

Figure \ref{MagneticBz149} confirms that a magnetowave is growing in
the lepton cloud. The wave length of the oscillations at the rear side
of the lepton cloud is comparable to an electron skin depth. The wave
vector forms a right angle with the cloud velocity vector, which is
typical for the waves that are driven by the filamentation instability 
\citep{Bret05}. 
Additional waves with a much shorter wave length are present within
and outside of the cloud. These waves arise from a finite grid
instability and they are thus a numerical artifact. Their frequency
exceeds by far the plasma frequency, which allows them to leave the
lepton cloud. A second consequence of their high frequency $\omega \gg
\omega_p$ is that their interaction with the plasma is weak. The loss
of energy of the cloud to these waves is negligible in our
simulation. This instability and some of its consequences are
described elsewhere \citep{Dieckmann06,Godfrey13,Godfrey14}.

\begin{figure}
\centering
\includegraphics[width=\columnwidth]{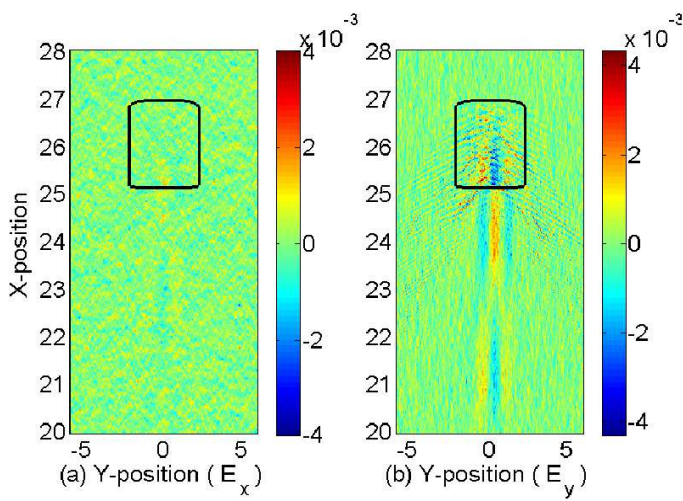}
\caption{The normalized in-plane electric field components at the time $t$=27.5 Panel (a) shows $E_x(x,y)$ and panel (b) shows $E_y (x,y)$. The color scale is linear. The black curve outlines the contour where the lepton cloud's number density is 0.06 $n_0$.}\label{ElectricExEy149}
\end{figure}

The in-plane electric field is displayed in
Fig. \ref{ElectricExEy149}.
The $E_x$ field remains at noise levels. The spatial distribution of
the electric $E_y$-component shows the same oscillations with a short
wave length, which we have already observed in
Fig. \ref{MagneticBz149}. Strong electric field oscillations of $E_y$
inside the lepton cloud and along the y-direction follow those of the
$B_z$-component in Fig. \ref{MagneticBz149}.

The direct comparison of the distributions of $B_z$ in
Fig. \ref{MagneticBz149} and $E_y$ in Fig. \ref{ElectricExEy149}(b)
reveals the source of the electric field. The electric field
distribution resembles the magnetic one, both qualitatively and
quantitatively. Their amplitudes match and this implies with our
normalization that $E_y (x_0,y_0) / B_z (x_0,y_0) \approx c$ at the
positions $(x_0,y_0)$ inside the region with the strong magnetic
field. The magnetic structure is confined to the lepton cloud and its
relativistic motion gives rise to a convective electric field in the
reference frame of the simulation box.

Figure \ref{ElectricExEy149}(b) shows electric field striations that
are aligned with the x-axis and trail the lepton cloud. The modulus of
the electric field amplitude of these striations is lower than that of
the convective electric field in the cloud and the phases of both
fields are shifted by $\pi$. These oscillations are tied to the return
current in the background electrons, which is induced by the
convective electric field in the cloud. This claim is supported by the
following estimate:

The amplitude modulus $|B_z|$ in Fig. \ref{MagneticBz149} reaches a
peak value of $B_{max} \approx 4.5 \times 10^{-4}$ and the thermal
velocity of the background electrons with a temperature $T_b = 1$ keV
is $v_{th} = 0.045$. The thermal gyroradius of the background
electrons would be $\approx 1000$, which exceeds the maximum extent of
the simulation box by an order of magnitude. The effect of the
magnetic field on the background electrons is thus negligible. The
background electrons are, however, exposed to the strong convective
electric field during a time interval $\delta_t = \delta_x / c$, where
$\delta_x \approx \lambda_e$ is the distance along x where the
convective electric field is strong. This time interval is thus
$\delta_t \approx \omega_p^{-1}$, which is long enough to enforce an
electron reaction. Ions with their much larger inertia would see the
convective electric field as a short impulse.

This bulk acceleration of the background electrons by the cloud's convective electric field generates a current, which induces an electric field in the background plasma. The convective and the induced electric field have opposite polarities, since the induced electric field tries to diminish the electric field of the driver. Once the cloud's convective electric field has left the perturbed plasma, the electrons will undergo high frequency oscillations at the plasma frequency $\omega_p$. High-frequency electrostatic oscillations in an unmagnetized plasma are known as Langmuir waves and they are only weakly damped; the oscillation will persist long after the cloud set it in motion. The electric field of the Langmuir oscillations will be polarized along the y-direction, since they were driven by the convective electric field with the same polarization direction.

The motion along $x$ of the spatially localized electric field driver, which is confined to the cloud, together with the oscillation in time of the electric field that has been induced in the background plasma, imply that the latter should be phase-modulated along $x$. The induced electric field oscillates once during $2\pi / \omega_p$ and the cloud propagates for a distance $2\pi c / \omega_p$ during this time. The spatial period of the modulation along $x$ should thus equal $2\pi \lambda_e$. We will use the term wakefield to describe the two-dimensional electric field distribution in the background plasma. We will next verify that a wakefield is present in the simulation data.

The wave length of the oscillations along the x-direction, which are trailing the lepton cloud in Fig. \ref{MagneticBz149}(b), is $\lambda_L \approx 2\pi \lambda_e$ or $k_L = \lambda_e^{-1}$. These waves fullfill the resonance condition $\omega_p / k_L \approx c$ thus we can identify them as Langmuir waves driven by the convective electric field of the cloud.

We can summarize our findings as follows: A magnetic field grows in the reference frame of the lepton cloud, however the plasma instability responsible for the wave growth is not the filamentation instability in its most basic form. The classic filamentation instability is driven by a magnetic deflection of two counter-streaming electron beams moving in opposite directions, which is only possible if both beams move relativistically in the rest frame of the magneto-wave and provide the strong spatially alternating current that is necessary to sustain the oscillatory magnetic field. The mean speed of the cloud's electrons and positrons with respect to the magnetic field vanishes, because the latter does not propagate in the cloud frame. The current contribution of these two species is thus negligible, implying that they can not participate in a magnetic instability that is driven by strong currents.

The instability is instead triggered by an interaction of the magneto-wave, stationary in the reference frame of the cloud, and the background plasma. Let us consider the background plasma in the rest frame of the cloud. The fast motion of the background electrons and the ions, which form a positive and immobile charge density distribution in the reference frame of the background plasma, implies that their partial currents are strong in the reference frame of the cloud. Initially, they cancel out each other. A magnetic perturbation caused by the ever-present electromagnetic noise in PIC simulations, which is stationary in the cloud's reference frame, deflects the relativistically moving background electrons. The deflected electrons do no longer move in unison with the positive charge background and a net current develops in the cloud frame. This net current re-inforces the magnetic perturbation, which leads to an instability. The magnetic deflection of the electrons in the cloud frame is equivalent to their electric deflection in the simulation frame, which is the source of the wakefield in Fig. \ref{ElectricExEy149}(b).

The spatial separation of the background electrons from the positive charge background also introduces a spatial charge in the cloud frame, which gives rise to a net electric field. The growing magnetic field yields via Amp\`ere's law a rotational electric field. Both electric field components add up to a net electric field in the cloud frame, which will separate its electrons and positrons. Figure \ref{NetCharge149} is evidence of the separation of both species.

\subsection{Time t=45}

\begin{figure}
\centering
\includegraphics[width=\columnwidth]{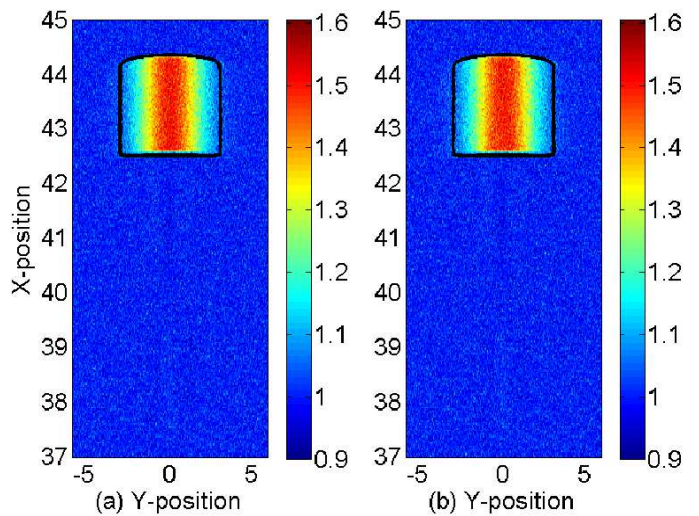}
\caption{The lepton density distribution at the time $t$=45. Panel (a) shows the normalized density $n_b (x,y) + n_e (x,y)$ of the background electrons and the cloud's electrons. Panel (b) shows the normalized density $n_b (x,y) + n_p (x,y)$ of the background electrons and the cloud's positrons. The color scale is linear. The black curve outlines the contour where the lepton cloud's number density is 0.06 $n_0$.}\label{Lepton246}
\end{figure}

Figure \ref{Lepton246} shows the number density distributions of the background electrons and the cloud's leptons at the time $t$=45.
The fastest leptons of the cloud have reached the position
$x\approx 44$. The lateral width of the cloud has increased further
due to the thermal motion of its electrons and positrons and the peak
densities have been reduced to a value of about $\approx 0.6$. The
electron and positron distributions look similar.

\begin{figure}
\centering
\includegraphics[width=\columnwidth]{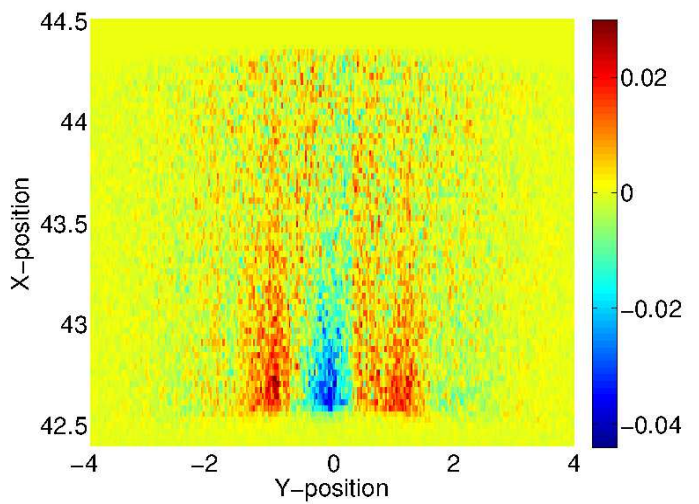}
\caption{The net charge density of the cloud at the time $t$=45. The net charge is calculated as $n_p (x,y) - n_e (x,y)$ (normalization to $n_0 e$) and the color scale is linear.}\label{NetCharge246}
\end{figure}

The number density difference $n_d (x,y)$ shown in Fig. \ref{NetCharge246} demonstrates that the spatial separation between the cloud's electrons and positrons has progressed, which implies that the instability is still growing.  
A pronounced minimum is observed at $y\approx 0$ and $x\approx 42.7$,
which is surrounded by two maxima. The amplitude of the oscillations
along the y-axis has more than doubled compared to those in
Fig. \ref{NetCharge149}, while the peak number density of the cloud
has decreased. The amplitude of the oscillations exceeds by far that
of the fluctuations.

\begin{figure}
\centering
\includegraphics[width=\columnwidth]{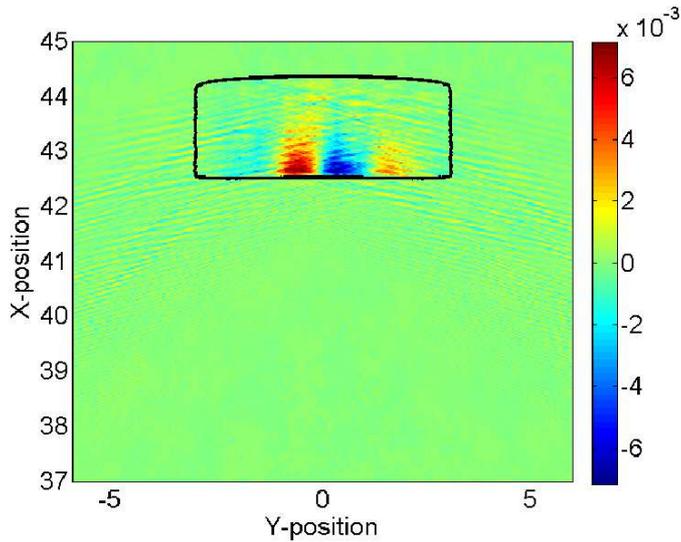}
\caption{The normalized magnetic amplitude $B_z (x,y)$ at the time $t$=45. The color scale is linear. The black curve outlines the contour where the lepton cloud's number density is 0.06 $n_0$.}\label{MagneticBz246}
\end{figure}

The amplitude of the magneto-wave in Fig. \ref{MagneticBz246} has
increased by almost 50 percent compared to that in
Fig. \ref{MagneticBz149}. Its wave length along the y-direction has
increased due to the thermal dispersion of the lepton cloud. The
instability appears to be robust against changes of the cloud
distribution; the reason being the briefness of the interaction
between the cloud and the background plasma. The steady stream of
unperturbed background electrons into the lepton cloud implies that
there are no memory effects in the plasma; the incoming electrons
simply adapt to the existing magnetic field structure and amplify it
with their current.  This scenario is not the case for the the
filamentation instability in an unbounded plasma. The particles are
heated up by the non-linear saturation of the latter, which results in
a destruction of its magnetic field structures. One reason is that hot
particles can no longer be confined magnetically to the current
channels that sustain the magnetic field. The current channels and the
magnetic fields they generate are dispersed by the thermal motion of
particles.
The magnetic field in Fig. \ref{MagneticBz246} is strong only inside
the lepton cloud. No magnetic field is present in the background
electrons behind the cloud.

\begin{figure}
\centering
\includegraphics[width=\columnwidth]{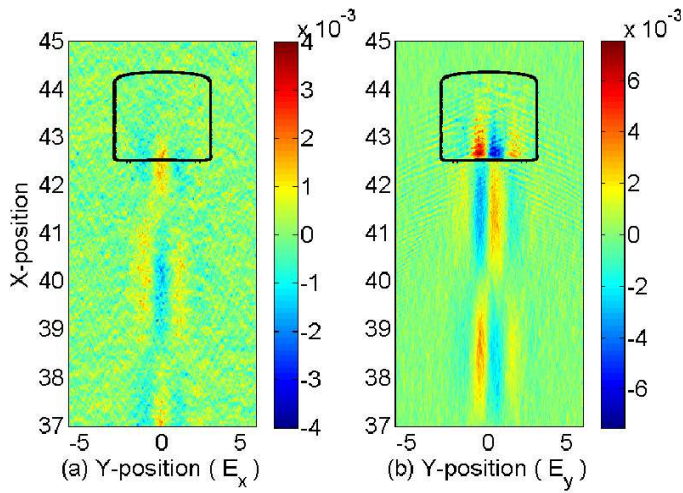}
\caption{The normalized in-plane electric field components at the time $t$=45. Panel (a) shows $E_x(x,y)$ and panel (b) shows $E_y (x,y)$. The color scale is linear. The black curve outlines the contour where the lepton cloud's number density is 0.06 $n_0$.}\label{ElectricExEy246}
\end{figure}

The larger magnetic field amplitude leads to a larger convective
electric field. Figure \ref{ElectricExEy246} reveals that this field
is now strong enough to induce visible wave oscillations in both
in-plane electric field components.
The wavelength of the oscillations of $E_x$ and $E_y$ are
correlated, shifted by $\pi/2$ in both in-plane
directions. The wavelength of the oscillations along the x-direction
is $2\pi$, while the wavelength along the y-direction matches that of the
electric field distribution inside the cloud. The trailing waves have
thus been driven resonantly by the convective electric field. No
corresponding structures are visible in the magnetic field, which
implies that the velocities of the background electrons are
non-relativistic. The convective electric field is not strong enough
to drive relativistic currents.

\subsection{Time t=110}

\begin{figure}
\centering
\includegraphics[width=\columnwidth]{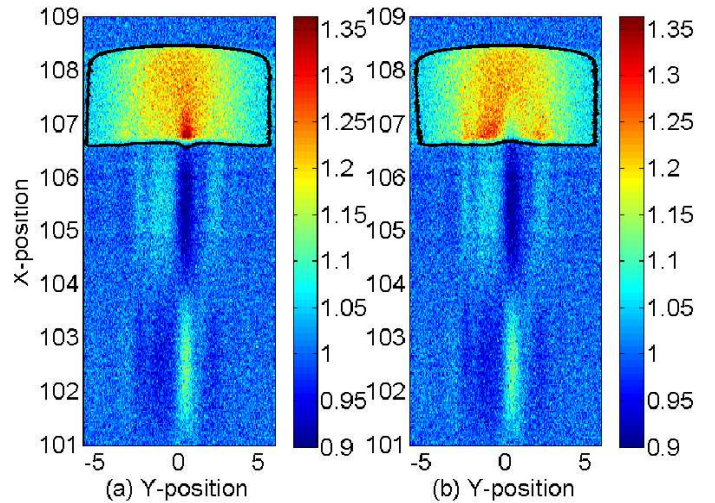}
\caption{The lepton density distribution at the time $t$=110. Panel (a) shows the normalized density $n_b (x,y) + n_e (x,y)$ of the background electrons and the cloud's electrons. Panel (b) shows the normalized density $n_b (x,y) + n_p (x,y)$ of the background electrons and the cloud's positrons. The color scale is linear. The black curve outlines the contour where the lepton cloud's number density is 0.06 $n_0$.}
\label{Lepton603}
\end{figure}

Figure \ref{Lepton603} shows the density distribution of the
background electrons and the distributions of the cloud's electrons
and positrons.
The front of the cloud has reached the position $\approx 108$ and its
longitudinal width has remained almost unchanged. The lateral width of
the cloud is now about 12, which exceeds its initial value by a factor
$\approx$ 6.5. The consequence of this spread is that the peak of the
unmodulated cloud density at the front of the cloud has decreased by a
factor four compared to its initial value. The density modulation at
the rear end of the lepton cloud is now clearly visible. The density
of the background electrons is also modulated and the oscillation
amplitude is a few percent of $n_b$.

\begin{figure}
\centering
\includegraphics[width=\columnwidth]{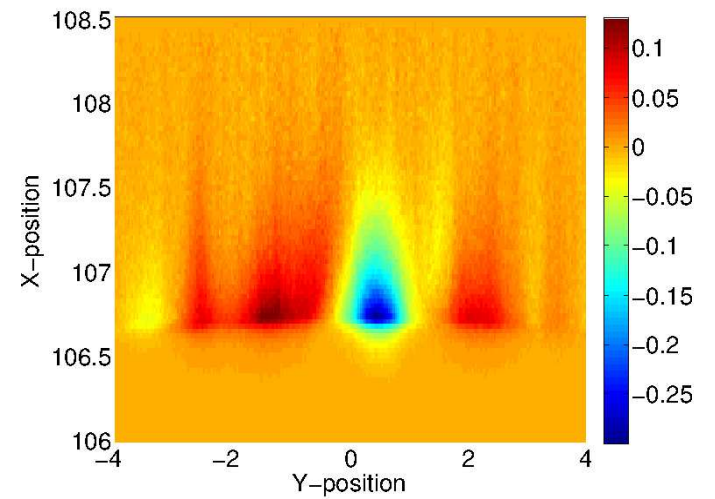}
\caption{The net charge density of the cloud at the time $t$=110. The net charge is calculated as $n_p (x,y) - n_e (x,y)$ (normalization to $n_0 e$) and the color scale is linear.}\label{NetCharge603}
\end{figure}

The number density difference $n_d(x,y)$, displayed in
Fig. \ref{NetCharge603}, now has a minimum value at $x\approx 0.5$ and
$y\approx 106.75$ that is comparable to the cloud's number density at
the front end.
The instability has reached its peak. From now on the thermal
dissipation of the cloud is likely to decrease the amplitude of the
density oscillation, at least in absolute terms.

\begin{figure}
\centering
\includegraphics[width=\columnwidth]{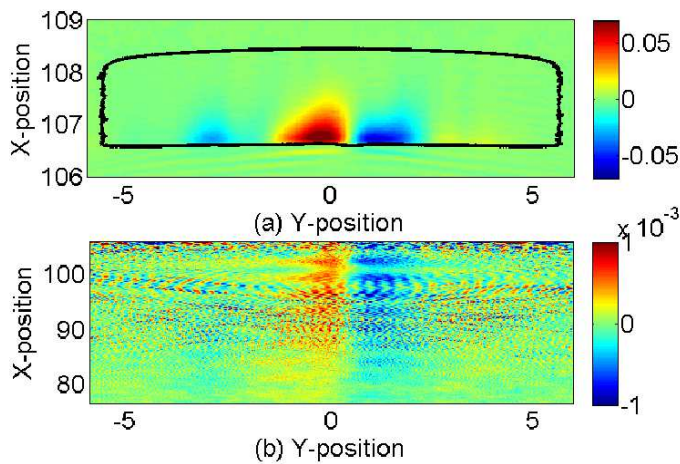}
\caption{The normalized magnetic amplitude $B_z (x,y)$ at the time $t$=110. Panel (a) shows the magnetic field distribution within the lepton cloud. The black curve outlines the contour where the lepton cloud's number density is 0.06 $n_0$. Panel (b) shows the magnetic field distribution behind the cloud. The color scale is linear.}\label{MagneticBz603}
\end{figure}

Figure \ref{MagneticBz603} shows that the magnetic field resulting
from the relativistically moving net charge of the cloud is
substantial. The peak value in Fig. \ref{MagneticBz603}(a) is $\approx
0.06$, which exceeds that at $t$=27.5 by more than an order of
magnitude.
A weak magnetic wakefield is revealed by Fig. \ref{MagneticBz603}. The
in-plane currents, which drive the electric field, also yield a weak
magnetic out-of-plane component. Its amplitude is, however, a factor
of 60 weaker than the magnetic field within the lepton cloud.

\begin{figure}
\centering
\includegraphics[width=\columnwidth]{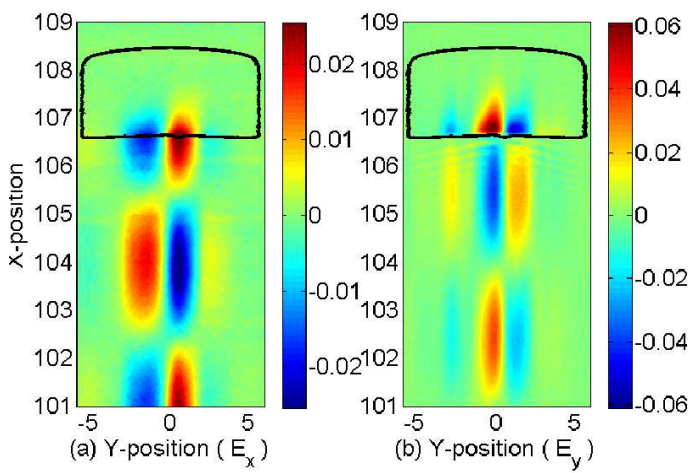}
\caption{The normalized in-plane electric field components at the time $t$=110. Panel (a) shows $E_x(x,y)$ and panel (b) shows $E_y (x,y)$. The color scale is linear. The black curve outlines the contour where the lepton cloud's number density is 0.06 $n_0$.}\label{ElectricExEy603}
\end{figure}

\begin{figure}
\centering
\includegraphics[width=\columnwidth]{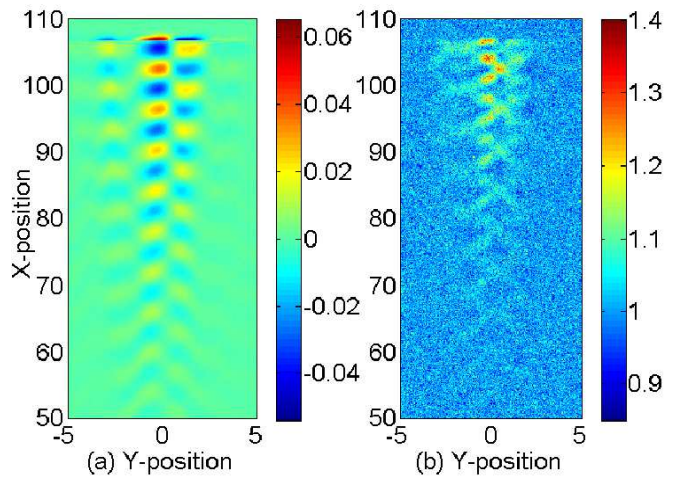}
\caption{Panel (a) shows the normalized electric $E_y$ component at the time $t$=110 over a long x-interval. Panel (b) displays the normalized energy density of the background electrons at the time $t$=110. The energy is expressed in units of their initial thermal energy density. The color scale is linear in both plots.}
\label{Energy603}
\end{figure}

Figure \ref{ElectricExEy603} reveals a strong electrostatic wakefield trailing the cloud. 
The peak amplitude value of these wakefield oscillations is about 0.03. The strength of this electrostatic wakefield becomes more evident when scaled to physical units. This electric field amplitude would be about 3 V/m in a plasma with an electron number density of 1 $\mathrm{cm}^{-3}$, which is typical for the interstellar medium. The energy density $K_e = \epsilon_0 \mathbf{E}^2 / 2$ corresponding to an electric field amplitude of 3 V/m, equals $10^4$ times the gas pressure $K_p = n_gk_B T_g$ of a gas with the density $n_g = 1 \mathrm{cm}^{-3}$ and temperature $T_g = 300$ K. The latter is representative for the interstellar medium far upstream of the jet's external shock. The peak value of the electric field would be 100 MV/m in a laboratory plasma with density $10^{15} \mathrm{cm}^{-3}$.  In spite of the observed large electric field amplitudes, the lepton cloud fails to magnetize the background plasma. A magnetic field is induced when the current $\propto v/c$ becomes more important than space charge, with respect to the generation of electromagnetic fields. The energy density distribution of the background electrons can provide insight into how far the background electrons are from reaching the relativistic speeds that will make the current strong. The latter are necessary to make the plasma dynamics magnetically dominated. Figure \ref{Energy603} shows the electric $E_y$ component and the kinetic energy density of the electrons over a wide spatial interval.
The electric wakefield in Fig. \ref{Energy603}(a) shows an amplitude
that increases with $x$. Langmuir waves have a group velocity that is
comparable to the electron's thermal speed at best. The group velocity
of Langmuir waves decreases with increasing wavelengths and we can
thus neglect wave propagation effects. The wakefield oscillations
remain localized and the electric field profile thus reflects the
temporal growth of the convective electric field that is moving with
the lepton cloud. The electric field becomes strong enough in the
interval $x>65$ to visibly enhance the kinetic energy density of the
electrons. However, the enhancement is limited to about 1.4 times the
initial value. The initial temperature of the background electrons has
been $T_b$ = 1 keV and the peak kinetic energy is thus far from being
relativistic.

\section{Discussion}

We have examined here the interaction of a tiny cloud of electrons and
positrons with a background plasma, via a particle-in-cell
simulation. The density of the cloud's electrons and positrons and
their mean speeds were set to be equal, thus the cloud was free of any
net charge and net current at the onset of the simulation. The lepton
cloud moved relative to a background plasma at a speed corresponding
to a Lorentz factor of 15. The background plasma consisted of mobile
electrons and a positive immobile charge background, and the density of
the background electrons matched the density of the cloud's electrons
in the simulation frame. All initial field components were set to
zero.

In our simulations we have seen an instability forming inside the
cloud, that in some ways is analogous to the filamentation instability
frequently invoked to explain the magnetic fields generation in the
jets of GRB's \citep{Medvedev99,Brainerd00}. The canonical
filamentation instability is driven by counterstreaming lepton beams
deflected into opposite directions by a magnetic perturbation. This
deflection groups leptons sharing the same direction of micro-current
vectors into channels; thus electrons flowing in the same direction
are grouped together, as are electrons and positrons flowing in
opposite directions.  The net current of these flow channels amplifies
the magnetic field.  This classical form of the filamentation
instability could not develop in our case study. The convection speed
of the magnetic field and the mean speed of the electrons and
positrons of the cloud all vanished in the reference frame of the
cloud. The instability we observe has thus been solely driven by the
interaction between the cloud magnetic field and the background
electrons, which are relativistic in the cloud frame.  This
instability is thus similar to the case of the magnetic barrier
considered by \citet{Smolsky96}. It can also be interpreted as an
extreme form of a filamentation instability between asymmetric lepton
beams \citep{Tzoufras07}.

The amplitude of the magnetic field inside the lepton cloud grew to a
value equivalent to 100 $\mu$G in an interstellar medium with an
electron number density of 1 $\mathrm{cm} ^{-3}$ or to about 6000 G in
a laboratory plasma with electron number density $10^{15}
\mathrm{cm}^{-3}$. The relativistic speed of the lepton cloud and the
magnetic field it carried created a strong convective electric field
in the reference frame of the background plasma. This field would be 
$10^{8}$ V/m in a laboratory plasma of density $10^{15} \mathrm{cm}^{-3}$ and
3 V/m in an interstellar medium with the density $1 \mathrm{cm}^{-3}$. The
energy density of this electric field would exceed the pressure of the 
interstellar medium by several orders of magnitude.
  
These electric fields were not strong enough to accelerate the background 
electrons to a speed that could induce significant magnetic fields in the 
background plasma, thus the magnetic field remained confined to the cloud. 
The electric field did however induce an electrostatic wakefield in the 
plasma. This wakefield constitutes an energy loss mechanism for 
relativistically moving pairs, able to slow down even the fastest pair 
flows, such as those emitted by AGNs. The simulation has shown that 
the magnetic fields remained high even though the thermal dispersion
of the lepton cloud implied that its shape and number density underwent
substantial changes. The lepton cloud can thus lose energy to the wakefield
over a long time, which leads to a substantial cumulative energy loss.

In this sense, the type of instability we observed in our simulation
could in principle also become important for the generation of
radiation by the jets of, e.g., GRBs, AGN and XRBs. For instance, it
has been suggested that some of the electromagnetic radiation of
relativistic jets is generated by the synchrotron jitter mechanism
\citep{Medvedev00,Keenan13}, where energetic leptons interact with a
small-scale magnetic field. The plasma close to collision-less shocks is non-thermal, which in a collision-less plasma can lead to phase space density distributions that are not uniform in space and that can have energetic beams. An accumulation of leptons that is spatially localized and confined to a small velocity interval could be interpreted as a spatially localized beam or a micro-cloud. In real systems, we expect non-thermal plasma throughout the jets where shocks are present. Compact lepton clouds escaping from the shocks could move independently and at moderately relativistic relative speeds through upstream plasma. Head-on
collisions between the upstream particles and a relativistically moving cloud that are mediated by the collective magnetic field may generate more energetic electromagnetic radiation than if these particles were to interact with a magnetic field that is stationary in the jet frame.

Yet another important consequence of the filamentation-type
instability we have observed here is that even microscopic lepton
clouds, which are charge- and current neutralized, undergo collective
interactions with the surrounding plasma. If this instability did not
develop, then the absence of binary collisions between plasma
particles would imply that these clouds could leave the vicinity of
the relativistic shock. The clouds would carry away energy and
constitute an energy loss mechanism. If this instability occurs in
real jets, it would ensure that these clouds cannot travel far and
hence they would deposit their mean flow energy into electromagnetic
fields and collisionless heating of the foreshock
plasma. The shock transition layer remains in this case spatially localized and its width does not increase steadily in time. The magnetic instability would contribute to a rapid plasma thermalization, which would make the shock behave like a fluid shock even if binary collisions between particles are negligible.

Our numerical results also serve as a motivation for further
laboratory experiments. The cloud size, the growth time of the
instability and the peak electric field are such that the instability
could be observed in a laboratory experiment.  These realizable scales
thus allow us to study plasma processes potentially giving rise to
some of the most violent outbreaks of electromagnetic radiation in a
controlled laboratory experiment.  Until now these could ony be
approached theoretically and by means of numerical simulations.

\textbf{Acknowledgements:} MED wants to thank Vetenskapsrådet for
financial support through the grant 2010-4063. GS and MB wish to thank
the EPSRC for supporting this work through the grants EP/L013975/1 and
EP/I029206/I. The Swedish High Performance Computing Center North
(HPC2N) has provided the computer time and support. SM is grateful to the University of Texas in Austin for its support, through a Beatrice Tinsley Centennial Visiting Professorship.

\bibliographystyle{aa}
\bibliography{Manuscript}

\end{document}